# HeM3D: <u>He</u>terogeneous Manycore Architecture Based on <u>M</u>onolithic <u>3D</u> Vertical Integration

AQEEB IQBAL ARKA[*], BIRESH KUMAR JOARDAR[*] RYAN GARY KIM[†], DAE HYUN KIM[*], JANARDHAN RAO DOPPA[*], AND PARTHA PRATIM PANDE[*]
[*]Washington State University, Pullman, WA
[†]Colorado State University, Fort Collins, CO

## ABSTRACT

Heterogeneous manycore architectures are the key to efficiently execute compute- and data-intensive applications. Through silicon via (TSV)-based 3D manycore system is a promising solution in this direction as it enables integration of disparate computing cores on a single system. Recent industry trends show the viability of 3D integration in real products (e.g., Intel Lakefield SoC Architecture, the AMD Radeon R9 Fury X graphics card, and Xilinx Virtex-7 2000T/H580T etc.). However, the achievable performance of conventional through-silicon-via (TSV)-based 3D systems is ultimately bottlenecked by the horizontal wires (wires in each planar die). Moreover, current TSV 3D architectures suffer from thermal limitations. Hence, TSV-based architectures do not realize the full potential of 3D integration. Monolithic 3D (M3D) integration, a breakthrough technology to achieve "More Moore and More Than Moore," and opens up the possibility of designing cores and associated network routers using multiple layers by utilizing monolithic inter-tier vias (MIVs) and hence, reducing the effective wire length. Compared to TSV-based 3D ICs, M3D offers the "true" benefits of vertical dimension for system integration: the size of an MIV used in M3D is over 100x smaller than a TSV. This dramatic reduction in via size and the resulting increase in density opens up numerous opportunities for design optimizations in 3D manycore systems: designers can use up to millions of MIVs for ultra-fine-grained 3D optimization, where individual cores and routers can be spread across multiple tiers for extreme power and performance optimization. In this work, we demonstrate how M3D-enabled vertical core and uncore elements offer significant performance and thermal improvements in manycore heterogeneous architectures compared to its TSV-based counterpart. To overcome the difficult optimization challenges due to the large design space and complex interactions among the heterogeneous components (CPU, GPU, Last Level Cache, etc.) in a M3D-based manycore chip, we leverage novel design-space exploration algorithms to trade-off different objectives. The proposed M3D-enabled heterogeneous architecture, called *HeM3D*, outperforms its state-of-the-art TSV-equivalent counterpart by up to 18.3% in execution time while being up to 19°C cooler.

## CCS CONCEPTS

• Computer systems organization ~ Architectures ~ Other architectures ~ Heterogeneous

## KEYWORDS

Heterogeneous manycore, NoC, M3D, performance, execution time, temperature, multi-tier

## 1 Introduction

Heterogeneous manycore systems that integrate multiple CPU and GPU cores on a single chip are widely used as enablers of data- and compute-intensive applications [1] [2]. Through silicon via (TSV)-based 3D ICs enable the design of high-performance and energy-efficient single chip heterogeneous systems [3]. However, the relatively large dimensions of TSVs (~um) present some fundamental limitations in high-performance, low-power manycore architecture design: (a) fine-grained partitioning of logic blocks across multiple tiers is not possible [4], forcing only planar implementations of cores and their associated logic elements; (b) Thick bonding material introduces heat dissipation challenges [5]; and (c) TSVs add non-negligible area and power overheads. Overall, TSV-based 3D designs cannot achieve the full-potential of vertical integration. This can lead to poor performance-thermal trade-offs in 3D CPU/GPU-based heterogeneous manycore design.

Meanwhile, monolithic 3D (M3D) has emerged as a promising technology for fine-grained vertical integration. In M3D, two or more tiers of devices are fabricated sequentially, one on top of another. This eliminates the need for any die alignment, which enables considerably smaller via sizes [4]. M3D integration uses nano-scale monolithic inter-tier vias (MIVs; with diameter of ~50nm) to connect the vertical device layers. MIVs are similar to regular metal-layer vias and can be used to design hardware logic over multiple tiers [4]. This results in significantly higher integration density than that of TSV-based 3D IC. In addition, this leads to better performance and energy efficiency. For instance, an M3D-enabled adder spanning two tiers outperforms conventional designs by 33% [6]. In a homogeneous manycore setting, M3D has been used to design high-performance network-on-chip (NoC) architectures that outperform its TSV-based counterpart by 28% [7]. However, a CPU/GPU-based heterogeneous manycore architecture using M3D integration is significantly more complex and remains unexplored.

A typical CPU/GPU-based manycore architecture principally consists of cores (CPU and GPU) that are connected using an NoC [8]. Designing the cores and the uncores (NoC and cache) using M3D reduces both area footprint and critical path length, thereby leading to better performance. Design of M3D-based CPU cores and caches have been proposed [9] [10]. However, to the best of our knowledge, a GPU design utilizing the benefits of M3D integration has not been undertaken. Moreover, the possibility of multi-tier logic blocks enables the design of high-performance and energy efficient NoC architectures [7]. The MIVs act as long-range shortcuts that bring processing elements (PEs) logically closer together, thereby leading to further performance improvements [7]. We utilize this feature to design a suitable high-performance NoC that augments the proposed CPU/GPU-based heterogeneous manycore architecture: *HeM3D*.

In addition, M3D provides better heat dissipation than TSV-based designs. Due to better thermal conductivity and extremely thin inter-layer dielectric (ILD) [5], heat is easily dissipated leading to reduced thermal hotspots. This is important as the dense circuit integration in 3D ICs gives rise to thermal hotspots that need to be addressed [5]. Furthermore, designing a 3D heterogeneous manycore architecture involves additional challenges: (a) the disparate nature of CPU and GPU architectures introduces conflicting design requirements that need to be satisfied simultaneously [8]; and (b) these architectures exhibit many-to-few-to-many communication patterns leading to on-chip communication bottlenecks [11]. Addressing these challenges is essential when designing a high-performance yet energy-efficient heterogeneous architecture.

Overall, M3D integration offers several performance benefits over TSV, which can significantly improve the performance of CPU/GPU-based heterogeneous manycore architectures. Our principal contributions in this work are:

1. We design an M3D-enabled vertical (3D) GPU core that outperforms conventional planar designs in terms of both performance and energy. The designed 3D GPU core is used in the proposed *HeM3D* architecture.

2. We present a design and optimization methodology that considers intrinsic M3D-related physical parameters to design and optimize the proposed *HeM3D* architecture.

3. We experimentally demonstrate that *HeM3D* simultaneously achieves better performance and lower temperature compared to its TSV-based counterpart for several well-known Rodinia [12] benchmarks.

The rest of the paper is organized as follows: Section 2 presents the relevant prior work. In Section 3, we discuss how M3D enables the design of various logic components over multiple tiers and present the overview of the proposed *HeM3D* architecture. Section 4 discusses the design and optimization strategy for the *HeM3D* architecture. Next, Section 5 presents the experimental results and analysis. Finally, in Section 6, we conclude the paper by summarizing the salient features of this work and discussing possible future research directions.

## 2 Prior Works

In this section, we present the relevant prior work for M3D-based integration and heterogeneous manycore architecture design.

### 2.1 Hardware Design Using M3D

TSV is the most popular integration methodology for 3D circuits and systems [3]. However, as discussed earlier, TSV-enabled architectures have several limitations, which affect performance, such as the lack of fine-grained partitioning and thermal bottlenecks [5]. Emerging M3D-based designs have the potential to address these challenges [4], [13]. In [13], the authors study the advantages of M3D integration by implementing transistor/gate-level partitioning and cell-on-cell stacking design. The speed and power benefits provided by M3D-enabled ICs have been investigated [14] [15].

In [16], the authors have presented the design of high-performance memory architectures using M3D. M3D-based high-speed cache was demonstrated to outperform state-of-the-art planar implementations [10]. A M3D-based CPU core with logic and memory partitioned in two tiers is demonstrated in [9]. High-performance M3D-based NoC design using multi-tier routers is explored in [7]. So far, the prior works have studied how M3D benefits individual components of a heterogeneous manycore architecture, namely cache, CPU, and NoC [7] [9] [10]. However, to the best of our knowledge, an M3D-enabled GPU design (similar to the 3D-CPU design of [9]) has not been undertaken previously and is necessary to design the *HeM3D* architecture. In addition, the performance gains of individual components do not translate fully to the overall performance benefits in a manycore architecture. Hence, a holistic design approach that simultaneously considers vertical core and uncore components in a heterogeneous manycore system is necessary.

### 2.2 CPU/GPU-Based Heterogeneous Architecture Design

CPU/GPU-based single-chip heterogeneous systems are well-suited for data- and compute-intensive applications [2]. However, due to their architectural differences, they tend to have conflicting design requirements, i.e., CPUs require low latency while GPUs demand high throughput [8]. This makes the design process challenging as these objectives need to be addressed simultaneously. In addition, CPU/GPU-based architectures exhibit an unbalanced many-to-few-to-many traffic pattern, which complicates the design further [11], [17]. Moreover, the power-hungry nature of GPU leads to thermal hotspots. which needs to be addressed as well [18], [19], [20].

A TSV-based 3D heterogeneous architecture can significantly boost performance but exacerbates the temperature problem. This happens due to the poor conductivity of the bonding material between planar layers [5], which does not allow heat to flow easily towards the heat sink. Prior works have tried to address these challenges in different ways [18], [21], [22]. However, as discussed earlier, M3D-enabled architectures have inherently better thermal profiles than their TSV-based counterparts. This property can be utilized to undertake aggressive performance optimizations that would be impossible using TSVs.

Altogether, in this work, we advance the state-of-the-art in heterogeneous manycore design by proposing *HeM3D*, that has the following features: (a) both core (CPU and GPU) and uncore (NoC and cache) components are designed vertically using M3D integration; (b) an optimized NoC that can handle the traffic hotspots caused by the many-to-few-to-many communication; (c) a machine-learning based multi-objective optimization (MOO) strategy that incorporates the features of M3D to quickly find suitable trade-offs between different design objectives; and (d) aggressive performance optimization to achieve high performance without creating thermal hotspots.

## 3 M3D Enabled HeM3D Design

In this section, we discuss the design of the M3D-enabled core (CPU, GPU) and uncore (NoC, cache, etc.) components of *HeM3D* and the overall *HeM3D* architecture. Figure 1(a) shows the *HeM3D* architecture where all CPU, GPU, and LLC tiles are multi-tier elements with logic and memory spread across two tiers with all tiles evenly placed across the four tiers. However, it should be noted that we only consider two-tier partitioning for the logic and memory blocks to simplify the design complexity and focus on the *HeM3D* architecture creation and optimization. To highlight the salient features of *HeM3D*, Figure 1(b) shows a regular TSV-based 3D heterogeneous manycore design (the equivalent of *HeM3D* using TSV). The TSV-enabled design (Figure 1(b)) utilize *planar* core and uncore components that are stacked on top of each other to create the 3D architecture. Hence, in a TSV-based design the main performance benefits arise from better network connectivity using the vertical links, not from improvements to the core and uncore elements.

M3D integration is enabled by fabricating two or more silicon layers sequentially on the same substrate and interconnecting the layers using small MIVs, allowing ultra-high-density fine-grained vertical integration [13]. This is fundamentally different from 3D integration using TSVs to interconnect separately fabricated dies [3]. Depending on the granularity with which devices are partitioned across multiple tiers, M3D-based architectures can be grouped into three main categories: (a) transistor-level or N/P partitioning: the nFETs and pFETs of a gate are placed on two separate tiers and connected via intra-gate MIVs [23]; (b) gate-level partitioning: planar gates are placed in different tiers and connected using inter-gate MIVs [24]; and (c) block-level partitioning: intellectual property (IP), functional, and memory blocks are placed in different tiers and connected using MIVs [25]. Among these different partitioning techniques, gate-level partitioning results in the highest amount of footprint reduction and subsequent performance improvement [24]. By placing different logic gates across multiple tiers, i.e., in 3D, the overall wirelength is reduced significantly. This leads to higher clock frequencies due to lower latency along the critical paths and a simplified and more energy-efficient clock tree and power delivery network [24]. Therefore, we adopt gate-level partitioning for *HeM3D* and discuss how each component in *HeM3D* benefits from using M3D next.

Figure 2(a) shows the planar cores used in TSV based heterogeneous manycore systems. Figure 2(b) illustrates the design of such cores using M3D-enabled gate-level partitioning in two tiers. These cores are paired with routers and the combination of a core and the associated router is referred to as a tile in Figure 1. The gates spread across different tiers are connected using MIVs. Here, the dimensions of the multi-tier tiles are considerably smaller than those of the planar tiles, so the critical paths of the multi-tier tiles are similarly shorter as well.

## 3.1 Vertical Core Design using M3D

*HeM3D* (Figure 1(a)) includes two types of cores: CPUs and GPUs. Next, we elaborate on the design of each type of core.

### 3.1.1 3D-CPU Design using M3D

M3D integration has been utilized to design high-performance yet energy-efficient 3D CPU cores [9]. The M3D CPU design in [9] is based on a typical pipelined, x86 architecture. By considering different stages of the CPU execution pipeline, we can identify critical paths and explore different design strategies to improve them. The various stages of the pipeline are then vertically partitioned across two tiers for best performance. Overall, compared to a conventional planar CPU, the M3D CPU improves critical path delay by 14%, which results in an average performance improvement of 14% and 26% in a single- and multi-core (4-cores) setting, respectively. We use the above-mentioned M3D CPU design [9] in the *HeM3D* architecture.

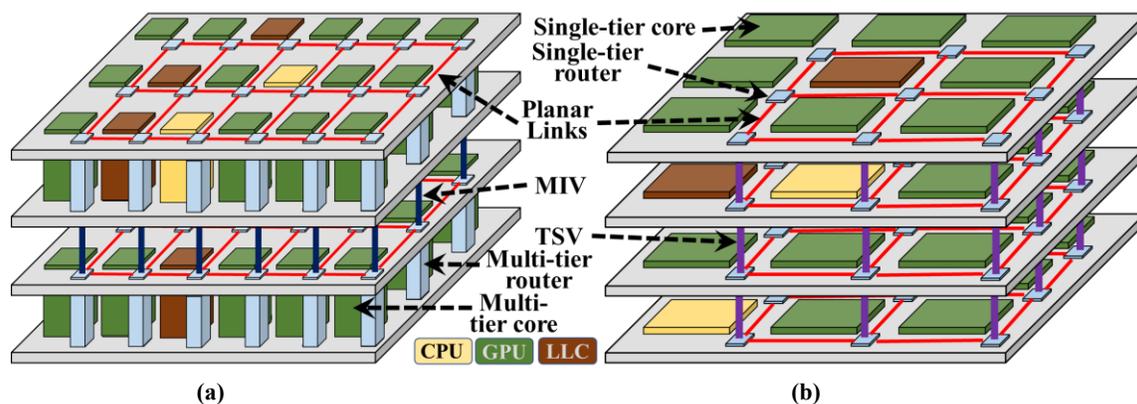

**Figure 1. Illustration of heterogeneous manycore architectures using (a) M3D (HeM3D) and (b) TSV. Tile and link placements are not optimized for any parameter, the figure is for illustration purposes only.**

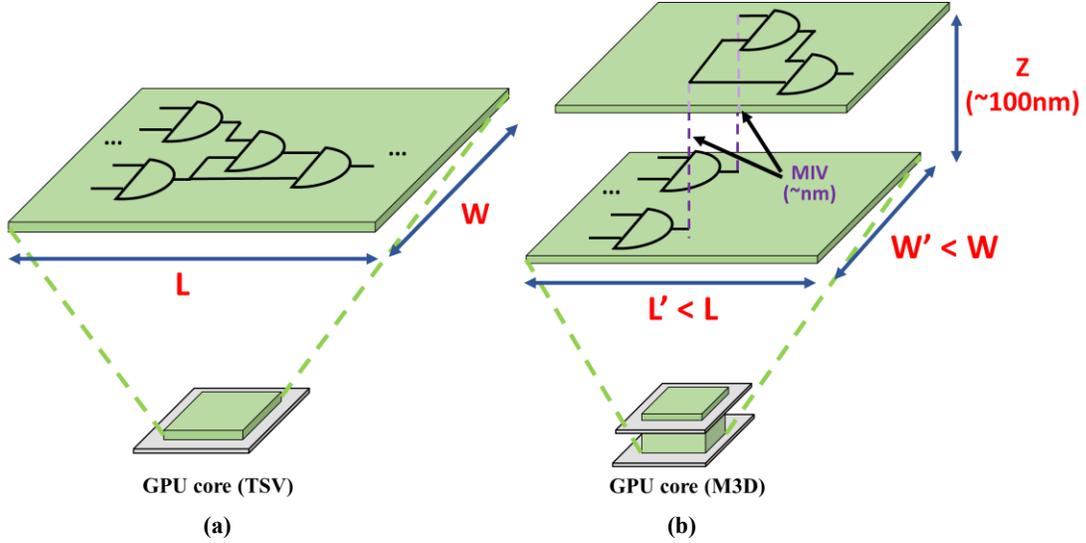

**Figure 2.** Illustration of (a) planar core (GPU is demonstrated as an example) and (b) multi-tier core enabled by M3D-based gate level partitioning. The width and length of the multi-tier core is substantially smaller than its planar counterpart. This is for illustration purpose only; it does not implement any specific logic.

### 3.1.2 3D-GPU Design using M3D

To the best of our knowledge, a GPU design using M3D has not been undertaken. It is well known that CPU and GPU architectures are fundamentally different. Therefore, it is not possible to predict the performance of an M3D-enabled GPU by following the M3D CPU design methodology. However, without M3D GPU cores, performance evaluation of the *HeM3D* architecture will be incomplete/speculative. Hence, it is imperative to first design an M3D GPU core.

A GPU core in *HeM3D* is analogous to a streaming multiprocessor (SM) in Nvidia terminology or a Compute Unit (CU) in AMD architecture. In this work, we use the open-source MIAOW GPU [26] for designing the M3D GPU. To the best of our knowledge, MIAOW is the only open-source RTL implementation available for a GPU core. However, please note that other GPU architectures can also be used to design the M3D GPU. The GPU core is made up of several blocks that operate in a pipelined fashion (as shown in Figure 3): (a) fetch, (b) wavepool, (c) decode, (d) issue, (e) execution blocks including scalar ALU, vector ALU (Single Instruction Multiple Data aka. SIMD and Single Instruction Multiple Floating-Point aka. SIMF), and load-store unit (LSU), and (f) registers (scalar and vector general purpose registers). Next, we discuss the design of the 3D GPU core using M3D integration.

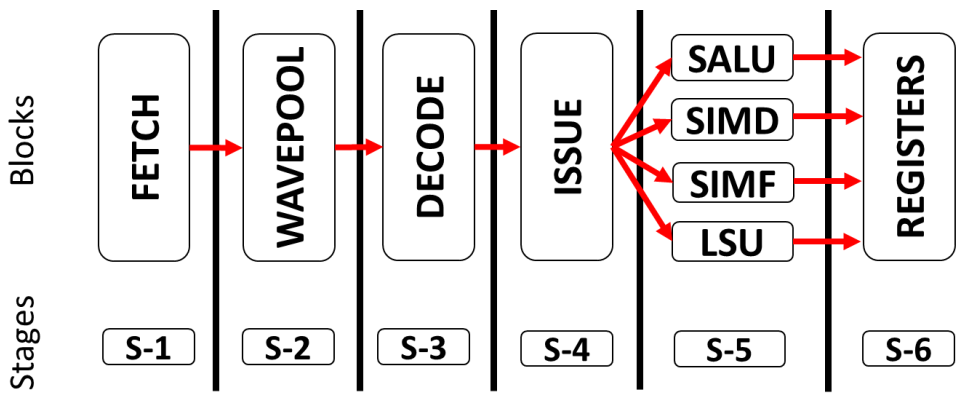

**Figure 3.** Basic execution pipeline of a GPU core.

**Designing the 3D-GPU:** To design the M3D GPU, we first synthesize a conventional planar GPU using Cadence Genus and a 45nm Nangate process. Then, we design (place, route, and optimize) a planar layout using the synthesized netlist and Cadence Innovus. Next, we use the planar GPU layout and the M3D IC performance prediction model proposed in [14] to get the performance of the M3D GPU. For this work, we made two modifications to the model proposed in [14] for placement and repeater optimization to showcase the potential of using M3D integration to improve performance of existing GPU architectures (a) Two consecutive inverters can be removed after 3D placement (uniform scaling of the instance locations) if the removal improves timing, and two consecutive inverters (without any branch point between them) can be considered a buffer and safely removed without altering the functionality of the netlist, and (b) The second modification is that a non-timing-critical branch with large load capacitance could be off-loaded from a timing-critical path by inserting a small buffer on the branch, thereby improving the timing characteristics of the path in the M3D design.

The M3D IC performance prediction model works as follows. For a given planar layout, it first scales down all of the gate locations by a uniform factor, reducing the length of each net ideally by $1/\sqrt{N_T}$, where $N_T$ is the number of tiers [14]. After the scaling, the model finds an ideal repeater insertion solution for each path to find a minimum delay for the path. As length of nets is reduced, a path that has repeaters in the planar layout has fewer repeaters or does not have any repeaters in the M3D design. Overall, the delay of a path in the M3D design decreases from $d_g + d_r + d_w$ to $d_g + d_r' + d_w'$ where $d_g$ is the sum of the delays of functional gates, $d_r$ and $d_r'$ are the sums of the repeater delays in the planar and 3D designs respectively, and $d_w$ and $d_w'$ are the sums of the net delays in the planar and 3D designs respectively. Here, please note that the delays of the functional gates remain the same because they are not affected by the M3D design due to gate level partitioning which designs individual gates in 2D. To increase the accuracy of the performance analysis, we obtain very detailed information such as the pin capacitance, wire capacitance, and wire length of the planar layout from Innovus and apply the model in [14] to obtain the critical path delay of the M3D GPU. Overall, this leads to significant improvement of the critical path delay, which enables us to operate the GPU core at higher frequencies compared to its planar counterpart. In addition, the use of MIVs and a smaller number of buffers leads to considerable amount of energy saving. Figure 2(a) and 2(b) show the structural difference between a conventional planar GPU core and the M3D GPU in *HeM3D*. We show the efficacy of the M3D GPU core in Section 5.

## 3.2 Vertical Uncore Design using M3D

Apart from the cores, *HeM3D* consists of uncore elements: cache and NoC, which also benefit from M3D integration. We discuss the design of these uncore components next.

### 3.2.1 3D-Cache Design using M3D

In addition to the many cores, *HeM3D* includes a few last level caches (LLC). Although smaller in number, these LLCs play a key role in determining the overall system performance. Memory accesses to and from LLCs (which also contain the main memory controllers) cause many-to-few-to-many traffic patterns that bottleneck the NoC [18]. A slow LLC response can lead to delays in LLC-bound data packets/requests, creating performance bottlenecks. CPU performance can be particularly affected due to CPU's higher latency sensitivity. Hence, faster caches are desirable in a CPU/GPU-based manycore architecture. The dense M3D integration allows high-performance cache designs as shown in [10]. By investigating different types of partitioning for caches, such as bank stacking, bit line partitioning and word line partitioning in two tiers, the cache architecture proposed in [10] achieves up to 23.3% reduction in access latency. For a single core system, the faster M3D cache is able to improve the overall performance by 9.9%. We use this high-performance 3D-Cache in *HeM3D*.

### 3.2.2 3D-NoC Design using M3D

In a manycore architecture like *HeM3D*, the NoC can have a large influence on the overall performance. The multi-hop nature of conventional NoCs e.g. mesh, introduces high communication latency, which can lead to performance bottlenecks. Hence, improving only the performance of the core and cache will not improve the performance of the whole heterogeneous manycore system. In addition, conventional NoCs are sub-optimal for the many-to-few-to-many traffic pattern observed in CPU-GPU based manycore architectures [11]. By adding few long-range links, small-world NoC (SWNoC) can significantly outperform mesh under many-to-few-to-many traffic patterns [18]. Moreover, M3D enables the design of (a) multi-tier routers which act as virtual long-range shortcuts in the SWNoC, and (b) physically shorter links due to the reduction of the physical dimensions of cores and caches as shown in Figure 1(a). Overall, an M3D-enabled SWNoC is a suitable choice as the communication backbone of *HeM3D*. We elaborate the design and optimization strategy for the NoC in more details in the next section.

## 3.3 Overall HeM3D Architecture

Altogether, *HeM3D* (Figure 1(a)) incorporates the M3D enabled core and uncore designs discussed so far. The CPU, GPU, and LLC tiles are extended vertically across two tiers in *HeM3D*. Each of these two-tier components are then distributed among four tiers to get the overall 3D structure as shown in Figure 1(a). Each tile is connected to an NoC router for communication. The routers are connected to each other via an optimized SWNoC topology. Overall, by vertically partitioning cores and uncores across multiple tiers in M3D, *HeM3D* lowers critical path delay leading to better performance. In addition, the vertical partitioning of cores and uncores result in lower power consumption due to the reduced wirelength and fewer repeaters. The improved power-efficiency results in an inherent reduction of on-chip temperature, which is otherwise a serious problem in TSV-based 3D designs. Hence, *HeM3D* achieves better performance and thermal characteristics than its TSV-equivalent as demonstrated later in Section 5.

## 4 Learning-based Optimization for HeM3D Design

In this section, we present a machine learning-based design and optimization strategy to establish suitable performance-thermal trade-offs in *HeM3D*.

## 4.1 Design Optimization in HeM3D

To design a suitable *HeM3D* architecture, we first profile the application characteristics ($f_{ij}(t)$), which is a function of time and represents the temporally varying communication frequency (the number of messages divided by the execution time in cycles) between tile *i* and tile *j*. This is done offline (only once) using detailed Gem5-GPU simulations by creating checkpoints [27] after specific intervals of the application execution. This divides the entire application into N smaller windows (periods of time) which allows us to observe and include dynamic application characteristics, in the *HeM3D* optimization process.

To optimize *HeM3D*, we consider the placement of tiles (CPUs, GPUs, and LLCs) within a grid and links between tiles, since they have a huge impact on the performance of *HeM3D*. Next, we discuss the different design objectives that need to be simultaneously satisfied to obtain an appropriate *HeM3D* architecture followed by the overall optimization process.

**CPU:** Due to their use of instruction-level parallelism, CPUs are latency sensitive. Hence, they need to be placed closer to the LLCs for faster access to main memory. For *C* CPUs and *M* LLCs, we approximate the latency of a *HeM3D* design *d* as:

$$Lat(d) = \underset{t}{avg} \left\{ \frac{1}{C*M} \sum_{i=1}^{C} \sum_{j=1}^{M} (r \cdot h_{ij} + d_{ij}) \cdot f_{ij}(t) \right\} \quad (1)$$

Where, $r$ is the number of router stages, $h_{ij}$ is the number of hops from tile *i* to tile *j* (CPU-LLC and vice versa) and $d_{ij}$ indicates the total link delay due to the physical separation (Euclidian distance based on the cartesian coordinates of the source and destination). Overall, we consider the average latency across all the $N$ number of $f_{ij}(t)$ instances of an application for the optimization purpose.

Multi-tier M3D routers change the hop count ($h_{ij}$) compared to a planar router [7]. The CPU-LLC latency in (1) reliably captures the effects of multi-tier routers enabled by M3D integration. M3D also allows partitioning of both core and uncore elements across multiple tiers. This reduces the physical dimension of each tile, hence, shorter wires, leading to lower effective link delay ($d_{ij}$) as shown in Figure 1. As a result, messages have to traverse less physical distance. Overall, $h_{ij}$ and $d_{ij}$, capture the effect of M3D in (1). In addition, as discussed in Section 3.1, M3D core and uncore components are faster than planar designs. This results in lower execution time, the effect of which is captured by $f_{ij}(t)$ in the form of lower execution cycles.

**GPU:** Unlike CPUs, GPUs rely on data-parallelism and require high throughput for high performance. Therefore, GPU placement and NoC connectivity should accommodate the huge volumes of data to/from the LLCs and the GPUs. The many-to-few-to-many traffic typically observed in CPU-GPU based architectures often lead to congestion in the NoC and poor performance. Hence, the *HeM3D* architecture should be optimized to achieve high throughput under such a traffic pattern. Load balancing is a popular way to address this problem [18]. Hence, we compare the degree of achievable load balancing in the NoC of designs ($d$) with different GPU and link placements using the mean $\bar{U}(d)$ and standard deviation $\sigma(d)$ of the traffic (load) on each link. These two metrics are calculated by the following equations, where the expected utilization of the k[th] link ($u_k$) is:

$$u_k = \sum_{i=1}^{C+M+G} \sum_{j=1}^{C+M+G} (f_{ij}(t) \cdot q_{ijk}) \quad (2)$$

$$\bar{U}(d,t) = \frac{1}{L} * \sum_{k=1}^{L} u_k \quad (3)$$

$$\sigma(d,t) = \sqrt{\frac{1}{L} \sum_{k=1}^{L} (u_k - \bar{U}(d,t))^2} \quad (4)$$

where $G$ represents the number of GPUs and $L$ indicates the number of links in the overall architecture. $q_{ijk}$ is a Boolean variable that indicates whether a link $k$ is used for communication between tile $i$ and tile $j$.

$$q_{ijk} = \begin{cases} 1, & \text{if cores } i,j \text{ communicate along link } k \\ 0, & \text{otherwise} \end{cases}$$

Here, it should be noted that the parameter $f_{ij}(t)$ includes the M3D effects during the GPU placement optimization by incorporating the reduced execution time, as discussed in the CPU section. Additionally, $f_{ij}(t)$ includes the effects of the many-to-few-to-many traffic. In order to represent the average throughput across all the time steps, we take a time average of the parameters represented by (3) and (4) above.

$$\bar{U}(d) = \underset{t}{avg}\ \bar{U}(d,t) \quad (5)$$

$$\sigma(d) = \underset{t}{avg}\ \sigma(d,t) \quad (6)$$

Minimizing the mean $\bar{U}(d)$ and standard deviation $\sigma(d)$ of the traffic distribution leads to overall higher throughput of the candidate *HeM3D* design.

**Temperature:** Next, it is well known that 3D integration has inherent temperature issues. Hence, it is important to consider the peak temperature ($T$) during optimization. We predict the maximum on-chip temperature using the following equation [19]:

$$T(d,t) = \underset{n,k}{\max}\{\sum_{i=1}^{k}(P_{n,i}(t) \sum_{j=1}^{i} R_j) + R_b \sum_{i=1}^{k} P_{n,i}(t)\} * T_H \quad (7)$$

Here, $P_{n,i}(t)$ is the power consumption of a tile $i$ tiers away from the sink in a vertical stack $n$ and is a function of application characteristics i.e. time dependent, $R_j$ is the vertical thermal resistance, $R_b$ is the thermal resistance of the base layer on which the dies are placed, and $k$ represents the $k^{th}$ tier where the tile is located. For an accurate prediction, we also consider the effects of lateral heat flow $T_H$, which represents the maximum temperature difference among all layers. The values of $R_j$ and $R_b$ depend on the material properties, which we obtain from [5] and calibrate using 3D-ICE simulations [20]. The worst-case temperature is considered during the optimization and is represented by the following equation:

$$T(d) = \max_t \{T(d,t)\} \quad (8)$$

Here, it is important to note that TSV and M3D systems have very different physical structures which affect temperature (illustrated in Figure 4). TSV-based architectures include a layer of bonding material between adjacent silicon tiers that has very poor thermal conductivity (Figure 4(a)) [5]. This prevents heat from easily flowing towards the heat sink. In addition, TSVs are much larger than MIVs, resulting in thicker silicon layers and a longer path for heat flowing towards the sink. Due to these reasons, a major portion of the heat spreads laterally rather than flowing vertically towards the sink (as shown in Figure 4(a)). As a result of this gradual heat accumulation among the layers, the overall temperature of the chip increases, which negatively affects the performance. Hence, TSV-based 3D integration is not very effective in designing high-performance heterogeneous architectures as we show later.

On the other hand, unlike their TSV counterparts, M3D integration (shown in Figure 4(b)) inherently exhibits better thermal properties due to thinner tiers and the absence of any bonding material [5]. The inter-layer dielectric (ILD) in M3D is significantly thinner and has better thermal characteristics than the equivalent "Bonding Layer" of TSV. This results in different effective thermal resistances for M3D- and TSV-based systems ($R_j$ and $R_b$ in (7)), which we obtain from [5] and 3D-ICE simulations [20]. In addition, the power consumption ($P_{n,i}$) in (7) also varies between TSV and M3D integration. M3D architectures are more power efficient than their TSV counterparts. Hence, by considering these M3D specific differences, we can accurately model the thermal profile in *HeM3D*. The lower temperature in M3D enables us to apply more aggressive performance optimizations to the overall architecture without having to worry about on-chip temperatures as we show later in this work.

**Overall:** Overall, we can formulate the CPU, GPU, LLC, and link placement problem to design the overall *HeM3D* architecture as a multi-objective optimization (MOO) problem. To highlight the different nature of trade-offs in TSV and M3D, we consider two flavors of optimization in this work: performance-only (PO) and joint performance-thermal optimization (PT). PO involves aggressively optimizing (placing links and tiles) for best performance, *i.e.*, low CPU-LLC latency and high GPU-LLC throughput under a many-to-few-to-many traffic pattern, without worrying about on-chip temperature. On the other hand, the PT optimization is more conservative and attempts to mitigate thermal hotspots while optimizing for performance. We show later that due to their physical differences, M3D- and TSV-based architectures require prioritizing one optimization strategy over the other. We can represent the two MOO-formulations as follows:

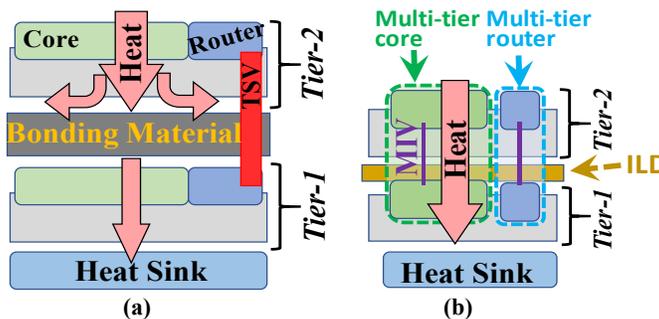

**Figure 4.** Illustration of physical structure and heat flow in (a) TSV two-tier cross-section, (b) M3D two-tier cross-section.

$$D^* = \begin{cases} MOO(OBJ = \{\overline{U}(d), \sigma(d), Lat(d)\}), & for\ PO \\ MOO(OBJ = \{\overline{U}(d), \sigma(d), Lat(d), T(d)\}), & for\ PT \end{cases} \quad (9)$$

where, $D^*$ is the set of Pareto optimal designs. Here, our aim is to find the optimal placement of both core and uncore elements such that the design requirements for all the elements are satisfied. We use an ML-enabled MOO solver, MOO-STAGE [18], that is shown to outperform conventional search and optimization algorithms. We elaborate the details of MOO-STAGE in the next section. Once we find $D^*$, we perform detailed full-system simulations using a cycle-accurate simulator Gem5-GPU [27] to obtain accurate energy consumption and execution times for each design. Then, we choose the solution which has the best EDP. We can formulate this using the following equations:

$$d^{best} = \begin{cases} argmin_{d \in D^*} ET(d), & , for\ PO \\ argmin_{d \in D^*} ET(d), such\ that\ Temp(d) < T_{th} & , for\ PT \end{cases} \quad (10)$$

Where, $\boldsymbol{d^{best}}$ is the chosen design that has the lowest execution time (ET) for PO or the lowest ET within a temperature constraint ($T_{th}$) for PT among all solutions in the Pareto front $\boldsymbol{D^*}$ and $\boldsymbol{Temp(d)}$ is the max temperature of the candidate design that is obtained using 3D-ICE [20]. For PT-based optimizations, $T_{th}$ was chosen to be 85°C so that it does not compromise the reliability of the manycore system [28], however, this can be chosen as per the designer's goals.

## 4.2 ML-enabled design Optimization

For manycore design and optimization problems, several algorithms including traditional Simulated Annealing-based AMOSA [29] and constrained combinatorial problems (CCPs) using SAT-decoding have been proposed [30]. However, SA-based algorithms like AMOSA, cannot explore large design spaces efficiently and often require significant amount of time to reach a good solution [18]. On the other hand, SAT-decoding based methods are not able to work with non-linear constraints (e.g. power-law based connectivity in SWNoC) and cannot be used to design *HeM3D*. Hence, there is a need for computationally-efficient design optimization algorithms.

For efficient exploration of such large design spaces, data-driven search techniques are necessary. Prior works including [18][31] have shown that by utilizing prior knowledge gained from exploring the search space, we can significantly reduce optimization time by focusing more on the promising regions of the design space. In this work, we employ MOO-STAGE, a MOO algorithm (introduced in [18]) which belongs to this class of data-driven search techniques, for the problem of M3D-enabled *HeM3D* design. Due to the use of dense multi-tier core and uncore elements, *HeM3D* design space is significantly larger than its TSV counterpart [7]. We show later that data-driven algorithms (e.g. MOO-STAGE) are far more effective in these scenarios compared to traditional algorithms.

Figure 5 illustrates the main steps of the MOO-STAGE algorithm. The key idea behind MOO-STAGE is to intelligently explore the search space such that the MOO problem is solved efficiently to uncover high-quality pareto sets. It utilizes a supervised learning approach that leverages past search experience (Local search) to learn an evaluation function. The evaluation function is then used to estimate the outcome of performing a local search from any given state in the design space (Meta search) to improve the accuracy of future searches in finding better solutions. As shown in Figure 5, MOO-STAGE is an iterative two-step algorithm: (a) Local search, and (b) Meta search. During the first step, MOO-STAGE performs a conventional search procedure (e.g., hill-climbing) from a given starting state guided by a heuristic function considering the different objectives to reach the local optima. For instance, in Figure 5, if the search begins at point *x11*, the local search process (assuming greedy heuristic) yields the local optima *x14*. Here, please note that hill climbing is used for the sake of simplicity only. More sophisticated search heuristics like IGOR [32] can also be used instead. The sequence of designs uncovered during past local search trajectories are used as training data for the next stage (aka. Meta search). The Meta search step is responsible for learning an evaluation function based on this input training data and attempts to predict the behavior of the search procedure from different starting states without actually executing it. The learned function predicts the potential (quantified using the quality of local optima, i.e., pareto hyper volume or the cost) of performing a local search starting from a given state. As shown in Figure 5, the learned function is able to correctly predict the region of the global minima (marked in green) based on the input training data from the past four searches. This allows the algorithm to discard bad starting states to reduce the number of local searches needed to find (near-) optimal designs in the given design space. In other words, the evaluator identifies the most promising areas for further evaluation over the next iterations. Here, it should be noted that, Figure 5 illustrates a snapshot in time after four iterations of MOO-STAGE. As the iterations (both Local and Meta Search) progress, the learned evaluation function changes dynamically and can predict the output of search procedure with higher accuracy as more training data becomes available.

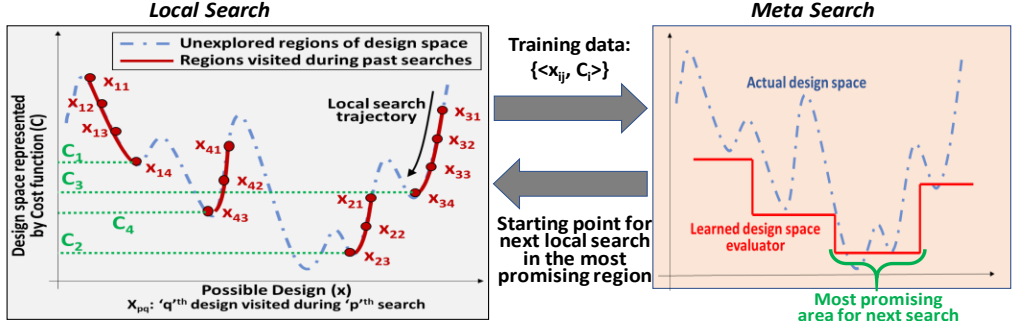

**Figure 5. Illustration of the two key steps of MOO-STAGE algorithm. Candidate designs (combinatorial structures) are represented using their cost values for ease of exposition.**

In this work, we use MOO-STAGE (after including the M3D-specific features discussed previously) as the MOO-solver for (9). In the next section, we experimentally show that AMOSA requires significantly more time to yield a good *HeM3D* design and is always outperformed by MOO-STAGE.

Algorithm 1 shows a high-level description of the design optimization strategy adopted in this work. Overall, given C CPUs, G GPUs, M LLCs and L links, the goal is to place all the tiles and links in a way that achieves best performance and thermal trade-off. For this purpose, we begin the design space exploration with a randomly chosen placement of the CPUs, GPUs, LLCs, and Links (referred as *D_curr*, Algorithm 1, Line 2). As mentioned earlier, MOO-STAGE is a two-step process repeated for MAX iterations (or till convergence). We begin the optimization algorithm with the local search (Algorithm 1, Line 4-7). In this work, we use greedy search for this step due to its deterministic nature, which is conducive to learning accurate evaluation function. Next, we perturb the candidate design (*D_curr*) to get a better new design (*D_next*, Algorithm 1, Line 5) quantified using a cost function. Here, a valid Perturb is defined as one of the following: (a) swapping the position of two tiles, or (b) Moving an existing link to a different source and destination pair. Each Perturb is followed by an evaluation phase that checks if the resulting design is valid based on a set of physical design constraints. For instance, the new design must have a valid path between any pair of source-destination tiles. This guarantees that resulting designs are practical. Each new solution (*D_next*) is characterized using a Cost which is a function of throughput, latency, and temperature of the new design (Algorithm 1, Line 5). In this work, we calculate the cost using the Pareto-Hyper Volume (PHV) metric as an example. Other cost functions such as weighted-average formulation can also be used here. The newly visited design is then stored in the local Pareto set based on general pareto dominance conditions (Algorithm 1, Line 7). In the next meta search phase of the MOO-STAGE algorithm (Algorithm 1, Line 8-12), the designs explored in the local search and their costs are used as training data to learn the evaluation function (Algorithm 1, Line 9-10). In this work, we use a regression tree learner (Algorithm 1, Line 10) for this purpose. Next, we randomly generate N valid solutions as potential candidate starting points for the next Local Search (Algorithm-1, Line 11). We use the learned model to predict the potential of yielding a good solution for each of these candidate starting designs (Algorithm 1, Line 12) without actually initiating a search process. The candidate with most potential (best predicted Cost) is used as starting point for the next Local Search iteration. The global Pareto set is updated using the local Pareto set after each MOO-STAGE iteration (Algorithm-1, Line 13). Finally, after max iterations (or after convergence) the designs from the global Pareto set, i.e., the set of non-dominated

| | |
|---|---|
| **Algorithm 1: HeM3D design using MOO-STAGE** | |

**Input**: Number of CPU, GPU, LLC tiles and planar links; Application characteristics ($f_{ij}$); Other design parameters like number of layers, how many tiles in each layer, etc.

**Output:** Optimized *HeM3D* design

**MOO-STAGE Algorithm:**

1    ***Initialization:*** *Training-set* [.]= {.}, *Pareto-Set* [.]={.},
     *D_curr* = Non-optimized HeM3D design
2    **Repeat till Convergence or MAX iterations**:
3      *Local-Pareto-set* [.] = {.}
4      **LOCAL SEARCH** *(Repeat till convergence):*
5        *D_next* = Best neighboring design from *D_curr* according to **Cost** function (e.g. PHV)
6        *D_curr = D_next*
7        ***Update*** *Local-Pareto-set* [.] using *D_next*
8      **META SEARCH**
9        **Add** new training data from LOCAL SEARCH (input designs, cost pairs) to *Training-set* [.]
10        *Model* = **Learn (***Training-set* [.]**)**
11        *D_random* [.] = *N* random valid designs
12        *D_curr* = **Predict** (*Model, D_random [.]*) with best **Cost**
13      ***Update*** *Pareto-Set* [.] using *Local-Pareto-set* [.]
14    **return** *Pareto-Set* [.]

solutions, are returned (Algorithm 1, Line 14). Detailed full-system simulations are then performed for each candidate design in this set to pick one solution for further evaluation. For the PO optimization, the solution with the lowest execution time is chosen as the final solution whereas, for the PT variant, the solution that has the lowest execution time within a temperature constraint ($T_{th}$) is considered for further analysis.

## 5 Results and Analysis

In this section, we first present the experimental setup to evaluate the *HeM3D* architecture. Next, we analyze the performance improvement of the M3D GPU core over its 2D counterpart. Then, we present the efficacy of the MOO-STAGE algorithm for optimization of *HeM3D* and finally, we evaluate the performance and thermal characteristics of the *HeM3D* architecture and compare it with a TSV-based baseline.

## 5.1 Experimental Setup

We utilize Gem5-GPU [27], a heterogeneous full system simulator to obtain processor- and NoC-level information. The CPU cores in *HeM3D* are based on Intel x86 while the GPUs are based on Nvidia Maxwell. As mentioned earlier, a GPU core in *HeM3D* is analogous to a streaming multiprocessor (SM) in Nvidia terminology. Each CPU and GPU core has a private L1 data and instruction cache of 32 KB. Each LLC consists of a shared 256KB slice of memory and a memory controller that provides access to main memory. To implement different TSV- and M3D-enabled NoCs used in this work, we modified the Garnet network [33] in Gem5-GPU. The memory system uses a MESI Two-Level cache coherence protocol. The planar CPU and GPU cores (baseline *core* architectures) operate at a frequency of 2.0GHz and 0.7GHz, respectively. The core power profiles have been extracted using GPUWattch [34] and McPat [35]. The corresponding on-chip temperatures have been obtained using the 3D-ICE simulator [20].

The TSV and M3D tiers are modeled in 3D-ICE using physical design parameters such as tier thickness, thermal conductivity of each tier, etc. as listed in [5]. The M3D cache behavior incorporated in our full-system simulations is modeled following [10]. The M3D-specific CPU-level improvements are obtained from [9] while GPU parameters are obtained from our synthesized M3D GPU (discussed in Sec III(B)). As mentioned in [9], the M3D-enabled CPUs can operate at 14% higher frequency (2.28 GHz) than its planar counterpart. We also obtained the M3D GPU core operating frequency to be 0.77 GHz (shown later) by following the design methodology elaborated in Section 3.2. Table 1 shows the relevant TSV and M3D parameters that we consider in this work. The inter-tier material (in Table-1) refers to the bonding layer in TSV-based integration and ILD for M3D-based integration (Figure 4).

Overall, *HeM3D* is a 64-tile architecture with eight x86 CPUs, 16 LLCs, 40 GPUs, and 64 routers (one per tile). The number of links in the SWNoC is the same as that of a mesh of same size. All TSV- and M3D-based architectures consist of four physical logic tiers. For the TSV-based architectures, the individual tiles (CPU, GPU, and LLC) are distributed over 4 planar tiers. On the other hand, M3D uses much smaller MIVs that enables vertical partitioning of logic blocks across multiple tiers (as discussed earlier). Hence, for *HeM3D*, we assume that each tile is spread across two tiers stacked on top of each other with equal area in both tiers as shown in Figure 1(a). Please note that the above architecture is only considered as an example. The proposed design and optimization methodology are generic and applicable for any system size/configuration. To evaluate the performance of *HeM3D*, we use six applications from the Rodinia benchmarks [12], namely, Backprop (BP), Needle (NW), Lava (LV), Lud (LUD), KNN, and Pathfinder (PF). We consider the application execution time obtained using full-system Gem5-GPU [27] simulations as the performance metric and the maximum on-chip temperature obtained from 3D-ICE [20] as the thermal metric.

## 5.2 Performance Analysis of 3D GPU Core

Before a full-system analysis of *HeM3D*, we first show the performance of the M3D GPU core (discussed in Section 3.2) and compare it with a traditional planar GPU core. For the baseline planar GPU core, we use Cadence Genus to synthesize the GPU using the Nangate 45nm process followed by a detailed placement and routing using Cadence Innovus. Next, we determine the timing characteristics of each block/pipeline stage of the GPU core. Figure 6 shows the timing characteristics of each pipeline stage normalized with respect to the clock period of the planar GPU core. Here, "S-i" denotes the "$i^{th}$" stage in the pipeline (Figure 3). S-1 being the first stage consisting of the "Fetch" block. In a pipelined architecture, the overall delay is bottlenecked by the slowest stage. From Figure 6, we can see that the pipeline stage delay (hence, the operating frequency) is limited by two stages: SIMD and LSU.

Next, we analyze the M3D GPU core designed using the methodology discussed in Section 3.2. As shown in Figure 6, M3D improves the timing characteristics of all the components in the GPU by 8% to 14%. However, as mentioned earlier, the slowest stage determines the clock frequency of the pipelined GPU. From Figure 6 we note that the SIMD stage has the highest delay in the M3D GPU design. Compared to the planar design, the M3D SIMD stage has 10% lower delay. Hence, we can operate the M3D GPU at 10% higher frequency without violating any timing constraints compared to its planar counterpart (baseline GPU core). This helps us to improve the M3D GPU core frequency to. 0.77 GHz from 0.7 GHz in planar implementation. In addition, we observe 21% lower energy consumption in the M3D GPU compared to its planar counterpart. We use this M3D GPU core in *HeM3D*.

## 5.3 MOO-STAGE enabled HeM3D design

As discussed in Section 4, the optimization of *HeM3D* is a MOO problem defined by (9) that can be solved using MOO-STAGE. Hence, in this section, we demonstrate the efficacy of the MOO-STAGE algorithm for the design and optimization of the *HeM3D* architecture. To evaluate the performance of MOO STAGE, we consider the well-known MOO algorithm AMOSA as the baseline due to its ability to achieve near optimal solutions [29]. We evaluate both algorithms based on their runtime and quality of solutions. Given the set of Pareto optimal solutions $D^*$, we run detailed simulations on each solution in $D^*$ to get accurate performance and temperature measurements. We use the same *HeM3D* MOO formulation (9) to design the TSV-based baseline architecture. In other words, we optimize the placement of CPUs, GPUs, LLCs, and planar links to improve latency, throughput, and temperature using TSV-specific parameters [18]. All experiments have been run on an Intel® Xeon® CPU E5-2630 @ 2.2GHz machine with 252 GB RAM running CentOS 6.

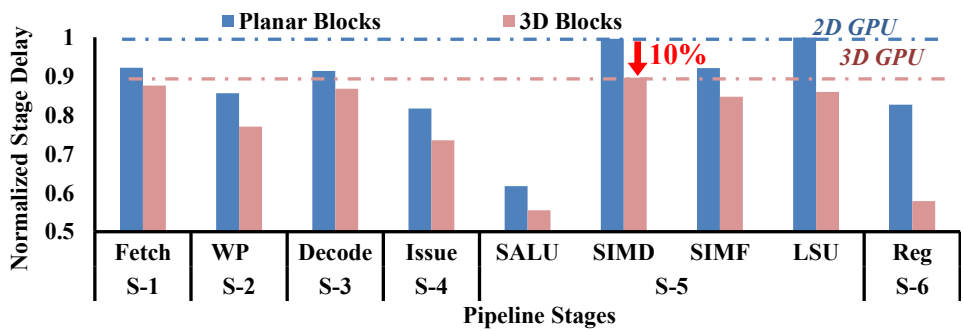

**Figure 6. Normalized pipeline stage latencies of the planar and M3D GPU core**

Figure 7 shows the speed-up in convergence time achieved by MOO-STAGE compared to AMOSA for designing *HeM3D* and its TSV-equivalent for all considered benchmarks. Here, we define convergence as the point of time beyond which the subsequent solutions vary in performance by less than 2%. This analysis is done with the joint performance-thermal optimization (PT) mentioned in (9). For the TSV-based design, MOO-STAGE outperforms AMOSA by 5.48x on average. This further increases to 7.38x on average for *HeM3D* design optimization. This happens as the design space of M3D-enabled architectures is significantly larger than their TSV-based counterpart [7]. Conventional search algorithms like AMOSA need to be annealed slowly and do not scale with the size of the search space. In addition, critical parameters like annealing temperature in AMOSA need to be tuned carefully for best results. Even then, AMOSA requires significant amount of time to yield a solution whose performance-thermal trade-off is comparable to that obtained using MOO-STAGE. On the other hand, by filtering out the promising regions of the *HeM3D* design space, MOO-STAGE virtually reduces the size of search space and hence the effort required to explore it to uncover (near-) optimal designs. As a result, MOO-STAGE achieves higher speed-up for *HeM3D* design.

## 5.4 Performance Evaluation of HeM3D

In this section, we analyze and compare the performance and thermal characteristics of *HeM3D* and its TSV-based counterpart. Figure 8 shows the performance and thermal trade-offs in TSV-based 3D heterogeneous manycore architectures. Here, we consider architectures designed following the two optimization strategies discussed in (9): (a) Performance optimized (PO): These architectures have the lowest execution time for a given application, and (b) Performance-Thermal joint optimized (PT): These architectures have the lowest execution time under a temperature threshold ($T_{th}$).

Figure 8(a) shows the *maximum* on-chip temperatures for both these variants. As shown in Figure 8(a), the on-chip temperature of the TSV-PO architecture can be as high as 105°C for some benchmarks. Such high temperatures are not desirable as it can affect the overall performance and reliability of the architecture. The poor thermal characteristics of the TSV-PO architectures arises due to a combination of the following factors: (a) The TSV-PO optimization places LLCs and most of the links in the middle two tiers [18]. This allows the LLCs to access the vertical links in both directions which reduces the average hop count to the other tiles. Also, the presence of more links enables greater path diversity, which improves load balancing under many-to-few-to-many traffic pattern (discussed earlier) leading to better performance. However, as a result, many power-hungry GPU cores get placed away from the sink [18]. (b) The layer of bonding material in TSVs has poor thermal conductivity (Figure 4(a)); preventing the heat to flow from these layers towards the sink; and

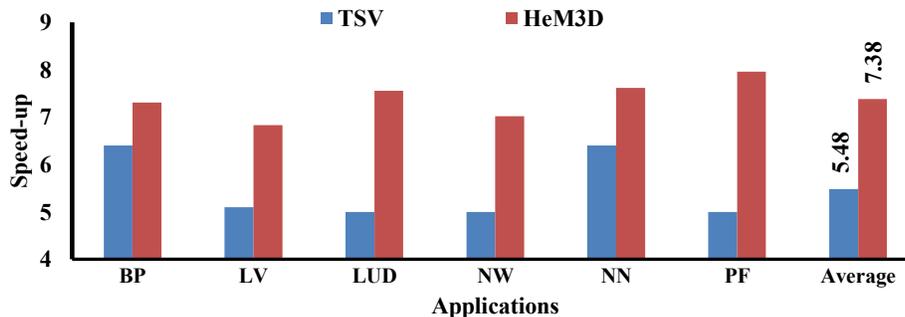

**Figure 7. Speed-up achieved by MOO-STAGE compared to AMOSA for designing *HeM3D* and its TSV-based counterpart**

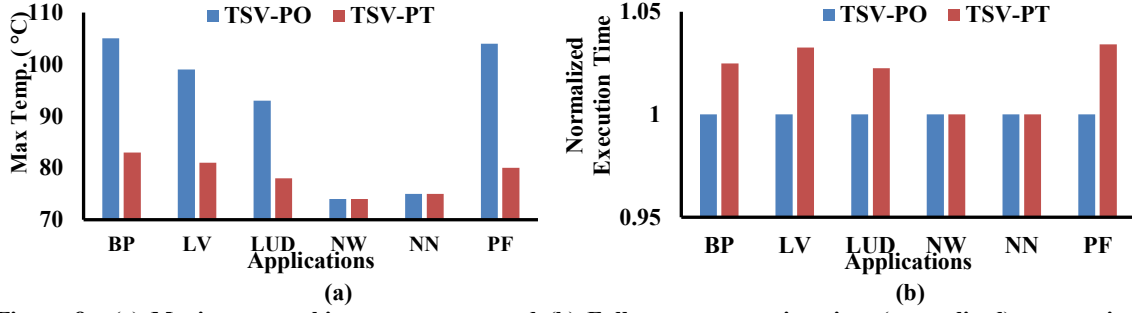

**Figure 8.** (a) Maximum on-chip temperature, and (b) Full system execution time (normalized), comparison between PO and PT optimized TSV-based architecture.

(c) the larger dimensions of TSVs result in lateral heat flow. This creates thermal hotspots as is evident from Figure 8(a). It is important to note that TSVs have better thermal conductivity than the bonding layer placed between two tiers. However, the total area occupied by the TSVs is much smaller than the area occupied by the bonding material. Thus, the improved thermal conductivity of TSV does not aid the heat flow [5].

On the other hand, TSV-PT lowers the temperature by up to 24 °C and 17.6 °C on average. This is due to TSV-PT's decision to place the power-hungry cores near the sink so that the generated heat can be easily dissipated. However, it should be noted that the temperature depends on the characteristics of the application. For applications with lower computational intensity (low *IPC*) like NW and NN, the temperature is relatively low. Hence, for NW and NN, the PT optimization ends up choosing the best performing option, in other words, the same design as PO. However, it is clear from Figure 8(a) that the PT-optimization is necessary for compute-intensive applications (BP, LV, LUD, and PF). Please note that microfluid-based cooling [20] has been used for both PO and PT optimized designs. Without liquid cooling, the temperatures in both cases become unmanageable. To achieve these lower temperatures, the PT optimization ends up sacrificing some performance as shown in Figure 8(b). The execution time for PT is 2-3.5% higher than its PO counterpart. However, this is a relatively small sacrifice to significantly reduce temperatures. This is important as higher temperatures are detrimental to the performance and long-term sustainability of the system. Thus, we can see that PT optimization with a thermal threshold constraint is necessary. In this work, we use TSV-PT as the baseline TSV architecture (TSV-BL).

On the other hand, M3D provides lucrative thermal advantages over its TSV-based counterpart, which we utilize to design *HeM3D* with better performance. Similar to TSVs, we explore both PO and PT optimization for *HeM3D*, *HeM3D*-PO, and *HeM3D*-PT, respectively, to study the performance-thermal trade-offs. Figure 9 shows the thermal (*maximum* on-chip temperatures) and performance characteristics of *HeM3D* for both PO and PT optimizations and compares it with the TSV baseline (TSV-BL) discussed earlier. As shown in Figure 9(a), both *HeM3D*-PO and *HeM3D*-PT have 18°C lower temperature than the TSV architecture on average. As shown in Figure 9(a), there is no temperature difference between *HeM3D*-PO and *HeM3D*-PT. Both architectures have temperatures in the range of 55-65°C for all the chosen benchmarks, which is much lower than the temperature threshold set for PT optimization. Due to this low maximum temperature, the *HeM3D*-PT optimization chooses the best performing option which is the same design as *HeM3D*-PO. The lower temperature of M3D compared to TSV is due to lower ILD thickness and better thermal conductivity. Since the ILD layers are extremely thin, virtually all the cores are near the sink. Hence, *HeM3D* achieves even lower temperature than TSV-PT, without the use of any specialized cooling. This lower temperature will make the *HeM3D* architecture more sustainable than its TSV-based counterpart.

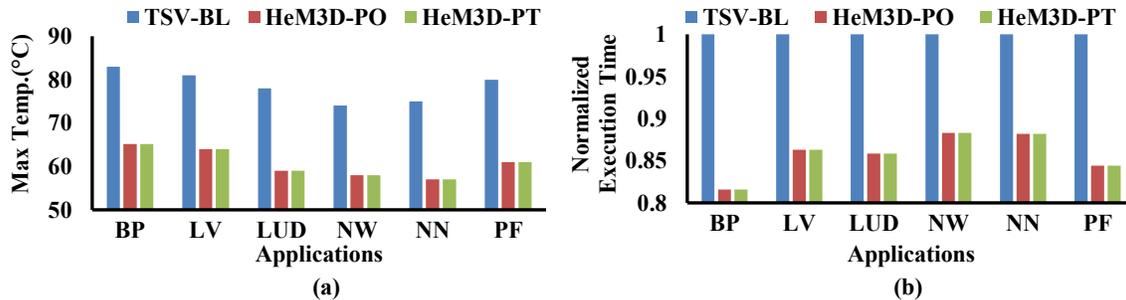

**Figure 9.** (a) Maximum on-chip temperature, and (b) Full system execution time (normalized), comparison between TSV-BL, *HeM3D*-PO. and *HeM3D*-PT.

Figure 9(b) demonstrates the normalized execution time for the three different architectures. From Figure 9(a), we note that *HeM3D*-PO and *HeM3D*-PT architectures outperform TSV-based counterparts by 14.2% on average. This happens as: (a) M3D cores operate at higher frequencies than planar cores (M3D CPU and GPU operate at 14% and 10% higher frequencies than their corresponding baseline planar counterparts); (b) the M3D-enabled cache provide 23.3% faster cache accesses (discussed in Section 3); and (c) the M3D routers and lower physical distance between adjacent cores (as shown in Figure 1(c)) enables high performance NoC designs which can easily handle many-to-few-to-many traffic.

However, instead of choosing the design with the lowest execution time under a temperature constraint set up by equation (10) for the *HeM3D*-PT architecture, if we choose the result with the lowest execution time-temperature product, we can improve the temperature of the architecture further as shown in Figure 10(a). Figure 10(a) illustrates that the *HeM3D*-PT design improves the temperature by 1-2 °C when compared to *HeM3D*-PO. On the other hand, Figure 10(b) shows that similar to TSVs, the *HeM3D*-PT sacrifices 2-3.5% performance compared to *HeM3D*-PO. From both Figures 9 and 10, we note that due to the unique features of M3D integration, PT optimization without thermal constraint does not bring significant benefits as it does for TSV designs [18]. *HeM3D*-PT loses up to 3.5% performance for a mere 1-2°C improvement in temperature. Hence, we can conclude that PT optimization is not necessary for *HeM3D*. Instead, M3D's unique thermal advantages can be utilized to push more aggressive performance optimizations which would otherwise be impossible in TSVs. Hence, we should use *HeM3D*-PO, which enhances performances by up to 18.3% while being 19°C cooler than a TSV-counterpart.

## 6 Conclusion and Future Research Directions

Emerging data- and compute-intensive applications require CPU/GPU-based heterogeneous manycore architectures. 3D integration is an emerging technology that has the capability to provide high-performance and energy-efficiency. However, existing TSV-based 3D architectures have thermal limitations and are insufficient to meet the performance and thermal requirements of these emerging applications. M3D enables the design of true 3D circuits and systems that mitigates many of the issues seen in TSVs. In this work, we utilize this feature to design a manycore heterogeneous architecture with vertical core and uncore elements: *HeM3D*. *HeM3D* outperforms TSV-based architectures in terms of both performance and temperature. More importantly, the superior thermal properties of M3D allows the use of a more aggressive performance optimization which is otherwise impossible with TSVs. Overall, the *HeM3D* architecture outperforms its TSV counterpart in terms of execution time by 18.3% while still being 19°C cooler.

However, it should be noted that M3D is an emerging technology with many unresolved challenges. Some notable examples include process variation [36][37] and electrostatic coupling among the different layers in M3D-based architectures [38]. As shown in [36],[37] and [38], these effects can lead to relatively sub-optimal performance for manycore architectures like *HeM3D*. A process-variation and electrostatic coupling aware design methodology can largely mitigate these negative effects and is the focus of our future work.

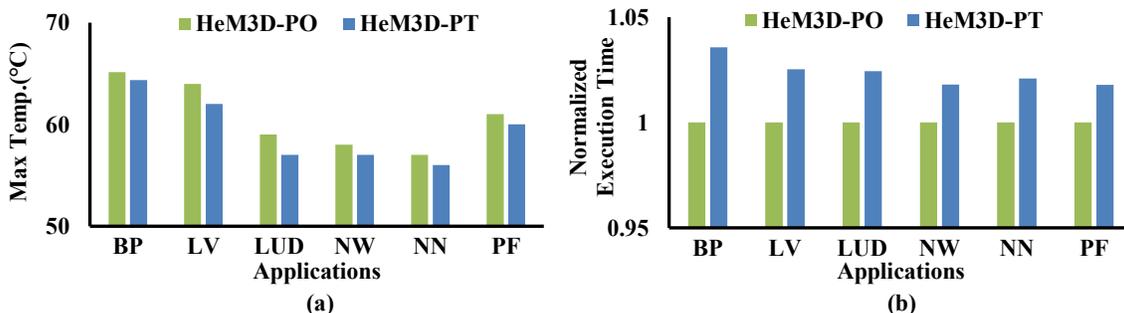

**Figure 10.** (a) Maximum on-chip temperature, and (b) Full system execution time (normalized), comparison between *HeM3D*-PO and *HeM3D*-PT without any thermal constraint.

## ACKNOWLEDGMENTS

This work was supported, in part by the US National Science Foundation (NSF) grants CNS-1955353, CNS-1564014 and USA Army Research Office grant W911NF-17-1-0485.